\documentclass[article]{aa}  

\usepackage{graphicx}
\usepackage{amsmath, bm}

\usepackage{txfonts}
\usepackage{hyperref}
\usepackage{natbib}
\bibpunct{(}{)}{;}{a}{}{,} 
\usepackage{color}
\usepackage{xspace}

\usepackage{listings}

\definecolor{codegreen}{rgb}{0,0.6,0}
\definecolor{codegray}{rgb}{0.5,0.5,0.5}
\definecolor{codepurple}{rgb}{0.58,0,0.82}
\definecolor{backcolour}{rgb}{0.95,0.95,0.92}

\lstdefinestyle{mystyle}{
    backgroundcolor=\color{backcolour},   
    commentstyle=\color{codegreen},
    keywordstyle=\color{magenta},
    numberstyle=\tiny\color{codegray},
    stringstyle=\color{codepurple},
    basicstyle=\ttfamily\footnotesize,
    breakatwhitespace=false,         
    breaklines=true,                 
    captionpos=b,                    
    keepspaces=true,                 
    numbers=left,                    
    numbersep=5pt,                  
    showspaces=false,                
    showstringspaces=false,
    showtabs=false,                  
    tabsize=2
}

\newcommand{\Gaia}{{\em Gaia}\xspace}

\begin{document}

\title{A machine learning-based tool for open cluster membership determination in \Gaia DR3}

\author{
    M.G.J. van Groeningen          \inst{\ref{inst:leiden}}\relax
\and   
    A. Castro-Ginard         \inst{\ref{inst:leiden}}\relax
\and   
    A.G.A. Brown          \inst{\ref{inst:leiden}}\relax
\and   
    L. Casamiquela          \inst{\ref{inst:paris}}\relax
\and   
    C. Jordi          \inst{\ref{inst:barcelona1},\ref{inst:barcelona2},\ref{inst:barcelona3}}\relax
}

\institute{Leiden Observatory, Leiden University, Niels Bohrweg 2, 2333 CA Leiden, the Netherlands\\ \email{mvgroeningen@strw.leidenuniv.nl} \relax \label{inst:leiden}\\
\email{acastro@strw.leidenuniv.nl} \relax \label{inst:leiden}
\and{GEPI, Observatoire de Paris, PSL Research University, CNRS, Sorbonne Paris Cité, 5 place Jules Janssen, 92190 Meudon, France \relax \label{inst:paris}}
\and{Departament de Física Quàntica i Astrofísica (FQA), Universitat de Barcelona (UB),  Martí i Franquès, 1, 08028 Barcelona, Spain \relax \label{inst:barcelona1}}
\and{Institut de Ciències del Cosmos (ICCUB), Universitat de Barcelona (UB), Martí i Franquès, 1, 08028 Barcelona, Spain \relax \label{inst:barcelona2}}
\and{Institut d'Estudis Espacials de Catalunya (IEEC), Gran Capità, 2-4, 08034 Barcelona, Spain \relax \label{inst:barcelona3}}
}

\date{Received date /
Accepted date}

\abstract{
        Membership studies characterising open clusters with {\em Gaia} data, most using DR2, are so far limited at magnitude $G = 18$ due to astrometric uncertainties at the faint end. 
}{
        Our goal is to extend current open cluster membership lists with faint members and to characterise the low-mass end, which members are important for many applications, in particular for ground-based spectroscopic surveys. 
}{
        We use a deep neural network architecture to learn the distribution of highly reliable open cluster member stars around known clusters. After that, we use the trained network to estimate new open cluster members based on their similarities in a high dimensional space, five-dimensional astrometry plus the three photometric bands.
}{
        Due to the improved astrometric precisions of {\em Gaia} DR3 with respect to DR2, we are able to homogeneously detect new faint member stars $(G > 18)$ for the known open cluster population.
}{
        Our methodology can provide extended membership lists for open clusters down to the limiting magnitude of {\em Gaia}, which will enable further studies to characterise the open cluster population, e.g. estimation of their masses, or their dynamics. These extended membership lists are also ideal target lists for forthcoming ground-based spectroscopic surveys.
}

\maketitle
%


\section{Introduction}
\label{sec:intro}

The study of open clusters (OCs) has gone through a rapid evolution in parallel with the different data releases of the {\em Gaia} mission \citep{Gaia_2016}. The major step forward was with {\em Gaia} DR2 \citep{2018A&A...616A...1G}, where the OC census was homogeneously studied for the first time taking advantage of the precise sky positions, parallaxes, proper motions and photometry in three different bands for more than one billion sources and the all-sky nature of {\em Gaia}. Using these data, \citet{2018A&A...618A..93C} were able to characterise over one thousand OCs in our Galaxy, providing accurate membership lists and mean astrometric parameters for them, and classify some objects present in pre-\textit{Gaia} catalogues \citep{Dias_2002,2013A&A...558A..53K} as asterisms. Moreover, the number of known OCs has increased with the discovery of hundreds of new objects, which only became detectable in light of {\em Gaia}. Assisted by novel machine learning techniques and a Big Data environment, \citet{Castro_Ginard_2018} systematically analysed the Galactic disc searching for new OCs based on the clustering of stars in the five-dimensional astrometric space, and then confirming them as real objects in {\em Gaia} photometry \citep{Castro_Ginard_2019,Castro_Ginard_2020,2022A&A...661A.118C}. Further studies contributed with new objects to the OC population to reach an OC catalogue which currently consists of around $2500$ objects \citep{2019JKAS...52..145S,2019ApJS..245...32L,2020MNRAS.496.2021F,2021A&A...646A.104H,2021MNRAS.504..356D}. For the whole OC catalogue, \citet{2020A&A...640A...1C} were able to estimate astrophysical parameters such as ages, distances and extinctions that enabled dynamical studies of this population \citep{2021A&A...647A..19T}, or the relation of the younger OCs with the spiral arms \citep{2021A&A...652A.162C,2021FrASS...8...62M}, providing a more complete view of the structure and evolution of our Milky Way.

All the previous studies rely on unsupervised learning techniques, mostly based on the clustering of stars, and have been limited to the bright end of the {\em Gaia} photometry, meaning stars with $G \leq 17$ or $18$ mag. Due to the increasing errors at fainter magnitudes, the compactness of the cluster is blurred and therefore the existing methodologies are less efficient in finding real OC members. This can be overcome by the inclusion of supervised learning techniques, able to learn the distribution of member stars around known OCs and find new members based on their similarities in a high dimensional space. This family of methods has already been applied to characterise stellar streams \citep{2011MNRAS.416..393B} or detect new ones \citep{2018MNRAS.477.4063M,2018MNRAS.474.4112M}, showing this is a powerful tool for this kind of objects (their elongated structure and wider range in parallaxes make them harder to study than OCs). 

Finding members to magnitudes fainter than $G = 18$ mag is important to fully characterise the OC population. The identification of low-mass members for OCs has many applications, to cite some examples, to test initial mass functions and mass segregation effects, investigate the limits between stars and planets or investigate the white dwarf population of these clusters. Having membership lists for the whole {\em Gaia} magnitude regime is also important for spectroscopic {\em Gaia} follow-up surveys. These forthcoming surveys, particularly WEAVE \citep{2012SPIE.8446E..0PD} and 4MOST \citep{2012SPIE.8446E..0TD}, are ground-based multi-object spectrographs that can observe around $1000$ and $2400$ objects simultaneously in a $2$ and $4$ squared degrees field-of-view, respectively. The target lists for both surveys are fully based on {\em Gaia} data and will complement {\em Gaia} with radial velocities and astrophysical parameters derived from spectroscopy for stars fainter than $G_{\text{RVS}} \sim 16$ mag, which is the {\em Gaia} spectrograph magnitude limit.  

This work takes advantage of the more precise astrometry and photometry of {\em Gaia} EDR3/DR3 \citep[][respectively]{gaia_edr3_survey_properties,2022arXiv220800211G}, with respect to DR2, to complement existing OC membership lists for bright magnitudes ($G \leq 18$) and find new members at the faint end. This paper is organised as follows. In Sect. \ref{sec:data} we show the steps for constructing a set of members, non-members and candidates for each cluster. In Sect. \ref{sec:method} we describe how we build a training and validation dataset with the members and non-members, how we train the neural network and how we apply the model to determine the membership probability of candidate members. To assess the performance of our method, we compare the OC membership lists we obtain with independently determined membership lists from \cite{Tarricq_2022} in Sect. \ref{sec:results}. Finally, we present our conclusions in Sect. \ref{sec:conclusions}.

\section{Data}
\label{sec:data}

We make use \Gaia DR3 \citep{2022arXiv220800211G} data to train our neural network to identify OC members. This data release contains astrometric (sky position, proper motion and parallax) and photometric (magnitudes in \Gaia's $G$, $G_{\mathrm{BP}}$ and $G_{\mathrm{RP}}$ bands) properties of more than 1.4 billion sources, which were first published in the previous data release: \Gaia EDR3 \citep{gaia_edr3_survey_properties}.

\subsection{Cone searches}
\label{sec:data_cone_searches}

For each OC we wish to study, a cone search is performed on \Gaia DR3 data to obtain data for sources in the sky vicinity of the OC. The cone search is centred on the mean sky position of the OC members, for which we use the values reported by \cite{2020A&A...640A...1C,2022A&A...661A.118C}. To determine the angular size of the cone search, we use an angular radius that corresponds to a projected physical radius of 50 pc at the location of the OC. This choice is based on the observation that OC cores are often surrounded by a halo or corona of comoving stars \citep{Meingast_2021, Tarricq_2022} which we want to include in our query. In addition, we only use sources within 10$\sigma$ from the cluster mean in proper motion and parallax space. The purpose of these cuts is to include both the most probable members and informative non-members in the cone search as well as minimising the computational load of the data processing.

\subsection{Members}
\label{sec:data_members}

We use \Gaia DR2 based membership lists assembled by \cite{2020A&A...640A...1C} to select the members which will be included in the training dataset. Most of these lists were collected from previous work by \cite{Cantat_Gaudin_2020_2, Castro_Ginard_2018, Castro_Ginard_2019, Castro_Ginard_2020} and some are the result of applying a clustering algorithm, UPMASK \citep{upmask_2014}, on OCs found by \cite{2019ApJS..245...32L}. For most OCs, these members only constitute the core of the cluster. We retrieve \Gaia DR3 measurements for these members by crossmatching their source identities with the corresponding cone search. For the training dataset, we only include members with a membership probability $p = 1.0$, which minimises the expected number of false positives among the members. The use of multiple OCs ensures a sufficient amount of members in the training dataset (see Sect. \ref{sec:training_set} for the construction of the training set).

\subsection{Candidate selection}
\label{sec:data_candidate_selection}

The sources in the cone search are then labelled as either candidates or non-members based on similarities to members of the corresponding OC in the dimensions of i) proper motion, ii) parallax and iii) magnitude and colour. For the proper motions, we consider as candidates the stars that satisfy
\begin{equation}\label{eq:pm_cond}
    \sqrt{\left(\frac{\mu_{\alpha^*} - \mu_{\alpha^*, \mathrm{c}}}{3\sigma_{\mu_{\alpha^*}} + \Delta_{\mu}}\right)^{2} + \left(\frac{\mu_{\delta} - \mu_{\delta, \mathrm{c}}}{3\sigma_{\mu_{\delta}} + \Delta_{\mu}}\right)^{2}} < 1,
\end{equation}
where $\mu_{\alpha^*}$ and $\mu_{\delta}$ are the proper motions of the star, $\sigma_{\mu_{\alpha^*}}$ and $\sigma_{\mu_{\delta}}$ are the uncertainties in the proper motions of the star and $\mu_{\alpha^*, \mathrm{c}}$ and $\mu_{\delta, \mathrm{c}}$ are the means of the proper motions of the OC members. The $\Delta_{\mu}$ is the maximum allowed separation in proper motion between candidates with negligible errors and the cluster mean. Conversely, $\Delta_{\mu}$ determines the minimum deviation for a source to be labelled a non-member. The value of $\Delta_{\mu}$ is different for each OC and depends on the sources we label as (training) members (Sect. \ref{sec:data_members}). How we determine the value of $\Delta_{\mu}$ is described in Sect. \ref{sec:maximum_separation_delta}.  The numerators in the fractions of Eq. \ref{eq:pm_cond} express a difference between the proper motion of a star and that of the mean of the cluster, whereas the denominators express a maximum deviation that candidates are allowed to have. Similarly, in parallax space, candidates must satisfy
\begin{equation}\label{eq:plx_cond}
    \left|\frac{\varpi - \varpi_{\mathrm{c}}}{3\sigma_{\varpi} + \Delta_{\varpi}}\right| < 1,
\end{equation}
with $\varpi$ the parallax of the star, $\sigma_{\varpi}$ its uncertainty, $\varpi_{\mathrm{c}}$ the mean parallax of the members and $\Delta_{\varpi}$ the maximum separation in parallax space. Finally, we select stars as candidates if they are close to the best fit theoretical isochrone \citep{2020A&A...640A...1C} of the OC
\begin{equation}\label{eq:iso_cond}
    \sqrt{\left(\frac{C - C_{\mathrm{ic}}}{3\sigma_{C} + \Delta_{C}}\right)^{2} + \left(\frac{G - G_{\mathrm{ic}}}{3\sigma_{G} + \Delta_{G}}\right)^{2}} < 1.
\end{equation}
where $C=G - G_{\mathrm{RP}}$ and $G$ are the colour and $G$ magnitude of the star, $\sigma_{C}$ and $\sigma_{G}$ are their uncertainties (derived with the tool provided by \Gaia DPAC\footnote{\url{https://www.cosmos.esa.int/web/gaia/dr3-software-tools}} to reproduce DR3 magnitude uncertainties), $\Delta_{C}$ and $\Delta_{G}$ are the maximum separations and $C_{\mathrm{ic}}$ and $G_{\mathrm{ic}}$ are the colour and magnitude of the isochrone point which is closest to the star. We use $G - G_{\mathrm{RP}}$ as the colour as \Gaia's $G_{\mathrm{BP}}$ band is known to overestimate the flux for faint sources, which causes the stellar distribution of an OC in the CMD to diverge from the isochrone \citep{Riello_2021}.

\begin{figure}
\centering
\includegraphics[width = 1.\columnwidth]{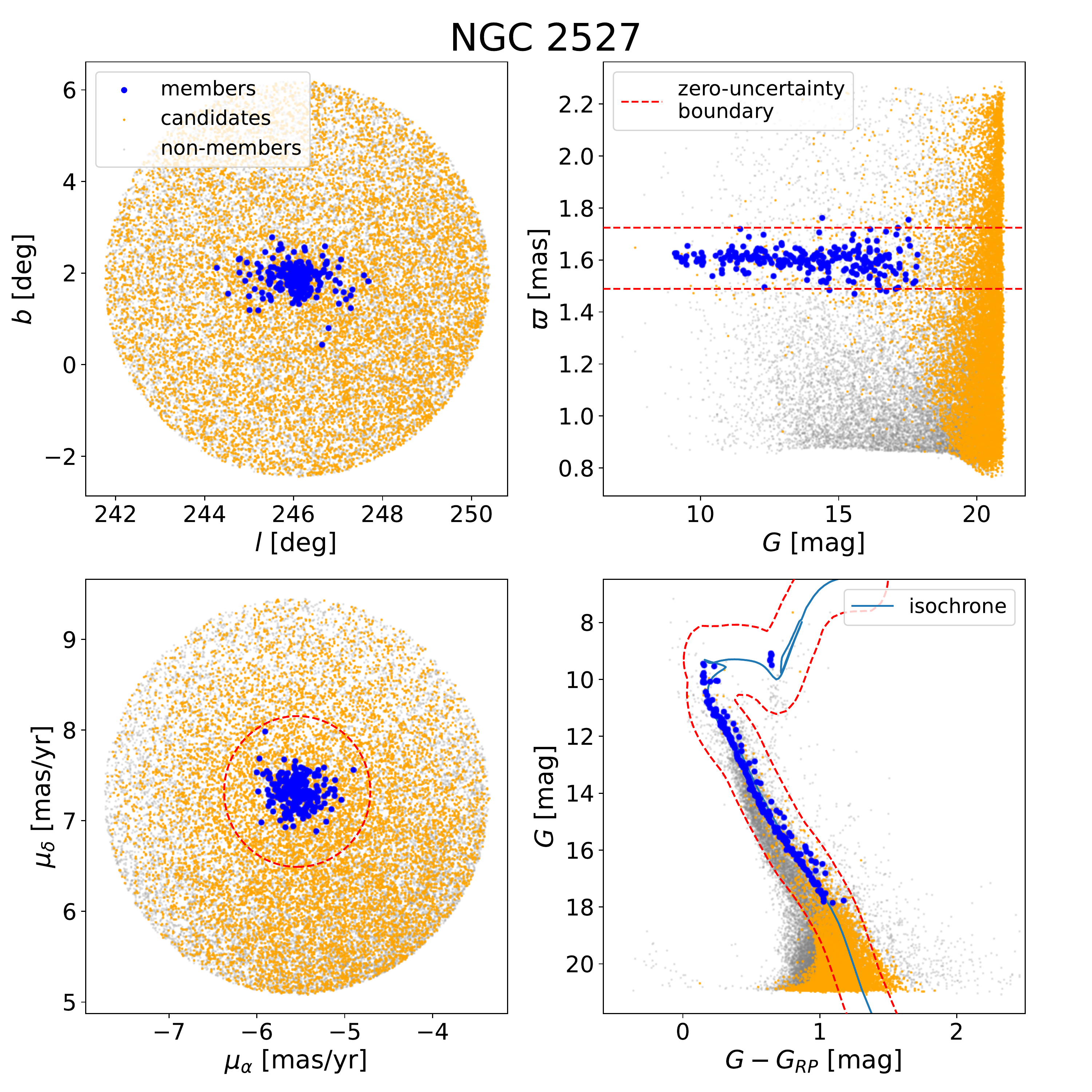}
\caption{Distribution of members (blue), candidates (orange) and non-members (grey) for NGC 2527 in sky position (top left), proper motion (bottom left), parallax (top right) and the CMD (bottom right). The blue line in the CMD constitutes the isochrone that corresponds to the age of NGC 2527 as provided by \cite{2020A&A...640A...1C}. The dashed red lines indicate a 'zero-uncertainty boundary', outside of which sources with negligible errors are not selected as candidates. Candidates that lay outside these boundaries thus have significant uncertainties.}
\label{fig:candidates}
\end{figure}

Candidates must then satisfy \textit{all} three conditions, such that they have both astrometric and photometric properties which are similar to those of the members. Figure~\ref{fig:candidates} shows the distribution of candidates selected by these conditions for the cluster NGC 2527. 

The isochrones, used for the CMD condition, are obtained through the Padova web interface\footnote{\url{http://stev.oapd.inaf.it/cmd}}, which computes the stellar evolutionary tracks with the PARSEC 1.2S and COLIBRI S37 models \cite{Bressan_2012},  \cite{Chen_2015}, \cite{Pastorelli_2020}, \cite{Marigo_2017}. To construct a compatible isochrone for each OC, we have used cluster ages, distances and extinctions reported by \cite{2020A&A...640A...1C} and adopted Solar metallicity. We correct the \Gaia magnitudes of the isochrone points for the cluster distance and interstellar extinction. To calculate the extinction for the $G$ and $G_{\mathrm{RP}}$ passband, we use a precomputed extinction model provided by the \texttt{dustapprox} Python package \cite{Fouesneau_dustapprox_2022}, which calculates the \Gaia band extinction for a given extinction $A_{0}$ at wavelength $\lambda = 550 \; \mathrm{nm}$.

\subsubsection{Maximum separation}
\label{sec:maximum_separation_delta}

As OCs are extended objects, the distribution of the members in the astrometric and photometric dimensions also depends on the morphology of the OC. To account for this feature in the candidate selection, we approximate the distribution in each dimension with a boundary which we parameterise with a maximum separation $\Delta$. The maximum separation $\Delta$ defines the maximum deviation from the cluster mean or isochrone that a source with zero uncertainties is allowed to have in order to be labelled as a candidate. In other words, it defines the boundary between candidates and non-members for sources with negligible uncertainties. This boundary is indicated by the red dashed line in Fig.~\ref{fig:candidates}. 

For the proper motion, we use
\begin{equation}\label{eq:pm_sep}
    \Delta_{\mu} = \sqrt{(3\sigma_{\mu_{\alpha^*, \mathrm{m}}} + 3 \sigma_{\mu_{\alpha^*, \mathrm{c}}})^{2} + (3\sigma_{\mu_{\delta, \mathrm{m}}} + 3 \sigma_{\mu_{\delta, \mathrm{c}}})^{2}},
\end{equation}
where $\sigma_{\mu_{\alpha^*, \mathrm{m}}}$ and $\sigma_{\mu_{\delta, \mathrm{m}}}$ are the standard deviation of the OC members in each proper motion component, while $\sigma_{\mu_{\alpha^*, \mathrm{c}}}$ and $\sigma_{\mu_{\delta, \mathrm{c}}}$ are the uncertainties of the weighted mean of the cluster proper motion components
\begin{equation}
    \sigma_{\mu_{i, c}} = \frac{1}{\sqrt{\sum_{j} 1 / \sigma_{\mu_{i, j}}}},
\end{equation}
where $\sigma_{\mu_{i, j}}$ is the error in the $i$-th proper motion component of the $j$-th member. For most OCs, the uncertainty in the cluster means is 10 to 100 times smaller than the standard deviation of the members, but for OCs with a small number of members which have relatively large errors, the uncertainty in the cluster means is significant.

For the parallax, we take into account the expected asymmetry in the parallax distribution, primarily for nearby OCs, due to the inverse relation between parallax and distance. We, therefore, use a different value for $\Delta_{\varpi}$ depending on whether the parallax of a source is greater or smaller than the cluster parallax
\begin{equation}\label{eq:plx_sep_cond}
\Delta_{\varpi} =\left\{\begin{array}{ll}
\Delta^{+}_{\varpi} & \text { if } \varpi<\varpi_{\mathrm{c}} \\
\Delta^{-}_{\varpi} & \text { if } \varpi\geq\varpi_{\mathrm{c}}
\end{array}\right.,
\end{equation}
where
\begin{equation}\label{eq:plx_sep_cond_+-}
    \Delta^{\pm}_{\varpi} = \left|\varpi_{\mathrm{c}}-  \frac{1000\; \mathrm{pc}}{\frac{1000 \; \mathrm{pc}}{\varpi_{\mathrm{c}}} \pm R_{\mathrm{max}}}\right| + 3\sigma_{\varpi_{\mathrm{c}}} + 3\sigma_{\varpi_0}.
\end{equation}
The first term in Eq. \ref{eq:plx_sep_cond_+-} is the difference between the cluster parallax and the parallax of a hypothetical source that lies $R_{\mathrm{max}}$ closer or farther away from the OC. We have used
\begin{equation}
    R_{\mathrm{max}}=R_{\mathrm{max}, 90} + 15 \; \mathrm{pc},
\end{equation}
where $R_{90}$ is the smallest projected radius to enclose 90\% of the members in sky position. The additional $15 \; \mathrm{pc}$ serves the purpose of a lower boundary for small OCs, while also taking into account that the training members generally only constitute the core of the cluster. The second term in Eq. \ref{eq:plx_sep_cond_+-}, parallel to the definition of $\Delta_{\mu_{i}}$, contains the uncertainty of the weighted mean parallax of the cluster. The third term contains an estimate of the uncertainty in the parallax zero-point $\varpi_{0}$, where we use $\sigma_{\varpi_0} = 0.015 \; \mathrm{mas}$ \citep{Lindegren_2021}, which is significant for distant OCs. We offset the parallaxes in our cone search with zero points as a function of magnitude, colour and ecliptic latitude according to the recipe provided by \cite{Lindegren_2021}.

Finally, for the colour and magnitude, we use

\begin{equation}\label{eq:delta_c}
    \Delta_{C} = \Delta_{C, 90} + 0.1
\end{equation}
and
\begin{equation}\label{eq:delta_g}
    \Delta_{G} = \Delta_{G, 90} + 0.8
\end{equation}
where we define $\Delta_{C, 90}$ and $\Delta_{G, 90}$ such that at least 90\% of our training members would pass the isochrone candidate condition (Eq. \ref{eq:iso_cond}) when $\Delta_{C} \geq \Delta_{C, 90}$ and $\Delta_{G} \geq \Delta_{G, 90}$. We additionally use the constraint $\Delta_{G, 90} /\Delta_{C, 90} = 8$ to obtain a single solution for each OC. This value approximately reflects the ratio between the ranges in colour and magnitude of sources in the CMD. By only letting 90\% of the members pass the isochrone candidate condition, we generally prevent $\Delta_{C}$ and $\Delta_{G}$ from being skewed by training members which do not follow the isochrone, e.g. blue stragglers. In contrast, the constant values added in Eq. \ref{eq:delta_c} and \ref{eq:delta_g} prevent the condition from being too restrictive, especially for OCs for which few of the training members deviate significantly from the isochrone.

\section{Method}
\label{sec:method}

In order to identify additional members of OCs, we make use of the Deep Sets (DS) neural network architecture developed by \citet{Zaheer_2017}. This architecture was designed to operate on sets, meaning unordered lists of objects, and therefore has the characteristic feature of returning the same output for every permutation of a given input. In our implementation of the DS architecture, we use this feature to perform the following classification task: given i) a set of stars which are labelled as members of the same OC (support set) and ii) an unlabelled candidate member for that OC, return a binary label, member or non-member, for the candidate. We train the neural network to recognise when a candidate star is sufficiently similar to the member stars in the support set, i.e. the members of the corresponding OC from \cite{2020A&A...640A...1C} with $p = 1.0$, in order to be classified as a member.

We use the same neural network architecture as \cite{Oladosu_2020}, who successfully applied the DS architecture to the analogous task of finding new members of stellar streams. They found that the DS architecture outperforms random forest baselines when trained and tested on synthetic data, i.e. a synthetic stellar stream inserted in a real field of stars extracted from \Gaia data, even when the random forest model was optimised for a subset of the members of the test stream in question. Compared to models that are trained on one specific stream, the DS architecture has the potential advantage of being able to learn higher-level member properties which are shared among streams. Another advantage, with respect to the random forest model, is that there is no need for negative examples (non-members) when applying the model to a new stream. However, when applied to one of the few actual stellar streams with reliable members, the fine-tuned random forest model did better than the DS architecture trained on synthetic streams, although a DS architecture optimised for the real stream performed best. \cite{Oladosu_2020} propose the difference in synthetic and real data as a possible explanation. In the case of OCs, thanks to recent developments in OC research \citep{Cantat_Gaudin_2020_2, 2019ApJS..245...32L, Castro_Ginard_2018, Castro_Ginard_2019, Castro_Ginard_2020}, we currently have the advantage of reliable membership lists for hundreds of OCs and can thus avoid the use of synthetic examples. In addition, the members of an OC generally follow a positional and proper motion distribution that is, for the majority of OCs, approximately spherically symmetric, which is easier to learn than the elongated structure followed by the stars in a stellar stream. Especially in parallax space, in which sources have relatively large uncertainties, the roughly similar distances of OC members pose less of a challenge than the gradient in distances of a stellar stream.

We have included diagrams of the model components in Appendix \ref{app:model_architecture}. For a more detailed description of the neural network architecture, we refer to \cite{Zaheer_2017}.

\subsection{Features}
\label{sec:training_features}

We attribute sources with a number of features on which the DS model has to base its membership predictions. For a feature to be effective, the (expected) distributions of members and non-members need to differ significantly in the feature space, as this enables the DS model to consistently differentiate the two classes. We use 5 source features, which relate to the sky position, proper motion, parallax, colour and magnitude of a source and 3 cluster features, which are the same for each source associated with a given OC. 

\subsubsection{Sky position separation}\label{sec:sky_pos_feat}
    
We use the projected radius $f_{R}$ between a source and the cluster centre
\begin{equation}\label{eq:radius_feat}
    f_{R} = D \cdot \theta,
\end{equation}
where $D$ is the distance to the OC with respect to us and $\theta$ is the angular separation between the source and the cluster centre, 
\begin{equation}
    \theta = \cos^{-1}\left[\sin(\delta)\sin(\delta_{c}) - \cos(\delta)\cos(\delta_{c})\cos(\alpha - \alpha_{c})\right],
\end{equation}

with $\alpha$ and $\delta$ the right ascension and declination of the source and $\alpha_{\mathrm{c}}$ and $\delta_{\mathrm{c}}$ the right ascension and declination of the cluster centre.

\subsubsection{Proper motion separation}

We use a 'proper motion separation' 
\begin{equation}\label{eq:pm_feat}
    f_{\mu} = \sqrt{\left(\mu_{\alpha^*} - \mu_{\alpha^*, \mathrm{c}}\right)^{2} + \left(\mu_{\delta} - \mu_{\delta, \mathrm{c}}\right)^{2}},
\end{equation}
which is a measure of a source's deviation from the mean proper motion of the OC. 

\subsubsection{Parallax separation}

Similar to the proper motion feature, we have
\begin{equation}\label{eq:plx_feat}
    f_{\varpi} = \varpi - \varpi_{\mathrm{c}}
\end{equation}
for a deviation measure in parallax space. 

\subsubsection{Isochrone vector}

The fourth and fifth features are the two components of a vector that represents a source's smallest separation from the isochrone
\begin{align*}
    f_{C} & = C - C_{\mathrm{ic}} \\
    f_{G} & = G - G_{\mathrm{ic}}
\end{align*}
where $[C_{\mathrm{ic}}, G_{\mathrm{ic}}]$ is the point on the isochrone for which
\begin{equation}\label{eq:iso_feat}
    d_{\mathrm{ic}} = \sqrt{\left(\frac{f_{C}}{\Delta_{C}}\right)^{2} + \left(\frac{f_{G}}{\Delta_{G}}\right)^{2}}
\end{equation}
is minimised. 

\subsubsection{Cluster features}

Besides the source-specific features, we also provide the model with a number of cluster features which are the same for each source to be classified for a given OC. These are the cluster mean parallax $\varpi_{c}$, the cluster age and the extinction of the cluster $A_{0}$.

\subsection{Training and validation set}
\label{sec:training_set}

A training and validation set are created from the members and non-members associated with 243 OCs. These OCs meet the following criteria: they (i) have their age, distance, extinction and at least 80 members with $p = 1.0$ available in the catalogue provided by \cite{2020A&A...640A...1C}, (ii) are not used to test the model (see Sect. \ref{sec:results}), (iii) have a Galactic longitude that deviates more than 60 degrees from the Galactic centre and (iv) have a parallax of less than $4$ mas. Conditions (iii) and (iv) aim to exclude OCs with computationally expensive cone searches. The validation set, which includes 30\% of these OCs, is used to monitor the performance of the model on unseen data during training. The remaining 70\% are contained in the training set and the performance of the model on this set determines the optimisation of the model parameters during the training process. By training on the members and non-members of many different OCs, the model is able to learn the general distribution of OC members, making it capable of finding new members even for OCs it has not been trained on.

Instances of both sets are created as follows: each member and non-member is first attributed with a number of training features (see Sect. \ref{sec:training_features}), which are designed to contain the relevant information of a source such that the model can make an accurate membership prediction. Next, we pair the member or non-member we want the model to classify with a support set, consisting of a random set of members (excluding the source to classify if it is also a member) of fixed size and from the same OC as the source to classify. We then combine the source to classify and the support set in a single tensor, which will be the input for the DS model. This tensor is created by concatenating the training features of the source to classify to the training features of \textit{each} member in the support set, resulting in a $N_{\mathrm{s}} \times 2M$ matrix where $N_{\mathrm{s}}$ is the number of members in the support set and $M$ is the number of training features per source. An instance of the training/validation set is then the pair of this input tensor and the binary label indicating whether the source to classify is a member or non-member.

In order to augment the number positive examples in our datasets, we create 2 instances with each member to classify for the training or validation set, depending on which set the corresponding OC is in. Both instances will contain the same member to classify, but a different random support set to prevent duplicity of the training/validation instances. From the set of non-members of each OC, we take 5 times the number of included members to classify (i.e. 10 times the number of unique members to classify) for that OC, which ensures a fixed ratio between members and non-members. The amount of non-members resulting from our candidate selection process is generally much larger than the number of members for a given OC and thus in most cases, all of the non-members to classify in the training and validation set are unique. In the case that the number of non-members we want to include for a given OC is larger than the number of unique non-members for that OC, we pad the difference with randomly selected non-members of that OC. 

\subsection{Training process}

\begin{figure}
\centering
\includegraphics[width = 1.\columnwidth]{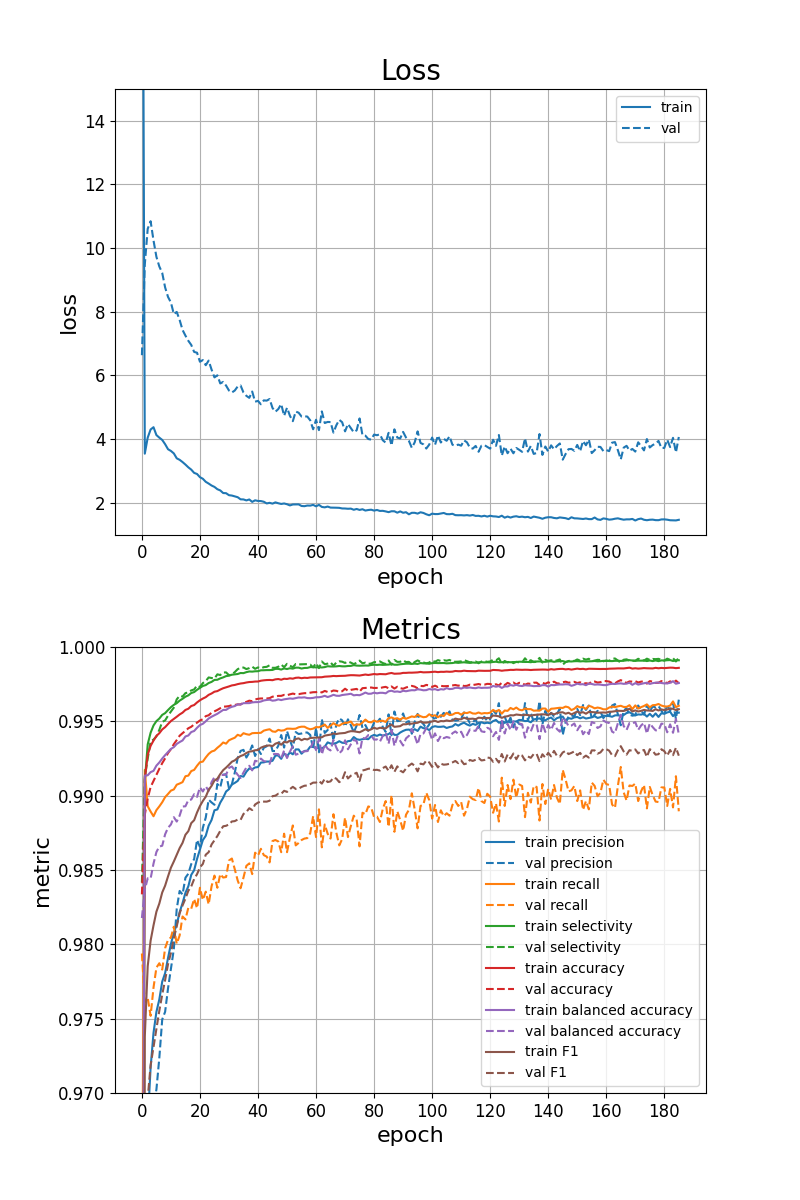}
\caption{Performance of the DS model during training. The top figure shows the evolution of the loss function (Eq. \ref{eq:loss}) for the training and validation set. The bottom figure shows the evolution of a number of classification metrics based on the number of true positives $TP$, true negatives $TN$, false positives $FP$ and false negatives $FN$, including: precision $=\frac{TP}{TP + FP}$, recall $=\frac{TP}{TP + FN}$, selectivity $=\frac{TN}{TN + FP}$, accuracy $=\frac{TP + TN}{TP + TN + FP + FN}$, balanced accuracy (average of recall and selectivity) and F1-score (harmonic mean of recall and precision). After 165 epochs, the model has reached its maximum validation F1-score.}
\label{fig:metrics}
\end{figure}

To optimise the model parameters, we use the cross-entropy loss function 
\begin{equation}
    L_{\mathrm{cross}} = - \sum_{i} \sum_{j} p_{ij} \log (q_{ij}),
\end{equation}
where $p_{ij}$ and $q_{ij}$ are, respectively, the true probability and predicted probability of source to classify $i$ and class $j$ (member or non-member). The true probability corresponds to the label of the source to classify and is therefore either 0 or 1. In order to mitigate overfitting, we apply two types of regularisation during training. We use L2 regularisation, giving a total loss function
\label{eq:loss}
\begin{equation}
    L = L_{\mathrm{cross}} + \gamma \sum_{i} w_{i}^{2},
\end{equation}
where $w_{i}$ are the trainable parameters and $\gamma$ determines the strength of the regularisation. In addition, we scale the gradients of the trainable parameters, which are used in the optimisation process, such that their norm does not exceed a certain value. We use \texttt{PyTorch}'s implementation of the \textit{ADAM} optimiser \citep{Kingma_2014} to minimise the loss function. To assess the performance of the model, we keep track of the F1-score, which is the harmonic mean of the recall and precision (see the caption of Fig. \ref{fig:metrics} for their definition). The F1-score is considered a suitable metric for data with a large class imbalance when the majority class is labelled as negative, which are the non-members in our case \citep{chicco2020advantages}. When the F1-score has not improved for 20 consecutive epochs, we stop the training process and use the model parameters that produced the maximum F1-score for the final model. Figure~\ref{fig:metrics} shows the evolution of the loss and a number of metrics for the training and validation set.

\subsection{Membership probability}
\label{sec:membership_probability}

We calculate a membership probability for each candidate member by applying the DS model on multiple samples of the candidate. For each sample, we re-calculate the proper motion, parallax, magnitude and colour of the candidate by sampling from a multivariate normal distribution defined by the candidate's uncertainties and the available correlations for these properties in the \Gaia data. With the sampled properties, we calculate the new training feature values of the sample. We also supply a different random support set for each sample. The membership probability is then defined as the fraction of samples for which the DS model identifies the candidate as a member. We use a sample size of 100 to cover both the variance in the feature values and the support set members.

\section{Results}
\label{sec:results}

The Python code and instructions for using the method are publicly available at \texttt{\url{https://github.com/MGJvanGroeningen/gaia_oc_amd}}.

To demonstrate the effectiveness of our method, we have tested the DS model on 167 OCs that (i) were provided with a membership list by \cite{Tarricq_2022} (T22) (ii) have their age, distance, extinction and at least 20 members with $p = 1.0$ available in the catalogue provided by \cite{2020A&A...640A...1C}, (iii) are not in the training or validation set and (iv) have a Galactic longitude that deviates more than 30 degrees from the Galactic centre to lighten the computational load. We compare the members we obtain to the members obtained by T22. T22 used the Hierarchical Density-Based Spatial Clustering of Applications with Noise (HDBSCAN) clustering algorithm \cite{Campello_2013}, which is considered a state-of-the-art method for determining OC members \citep{2021A&A...646A.104H}, to establish their membership lists. They ran HDBSCAN on {\em Gaia} EDR3 parallax and proper motion dimensions ($\varpi, \mu_{\alpha^*}, \mu_{\delta}$) and applied no additional selection criteria in sky position dimensions as T22 focused on studying the halos of OCs. Note that our method uses the same parallax and proper motion data as T22, but that it uses sky position and photometric data as well. 

We also considered comparing with membership lists from \cite{2021MNRAS.504..356D}, as they assembled membership lists from various sources and a significant fraction of these also include $G > 18$ members. However virtually all their OCs with $G > 18$ members do not have a membership list available in \cite{2020A&A...640A...1C}. As such, a systematic comparison in which only members from \cite{2020A&A...640A...1C} are used for the support set is not viable. 

\begin{figure}
\centering
\includegraphics[width = 1.\columnwidth]{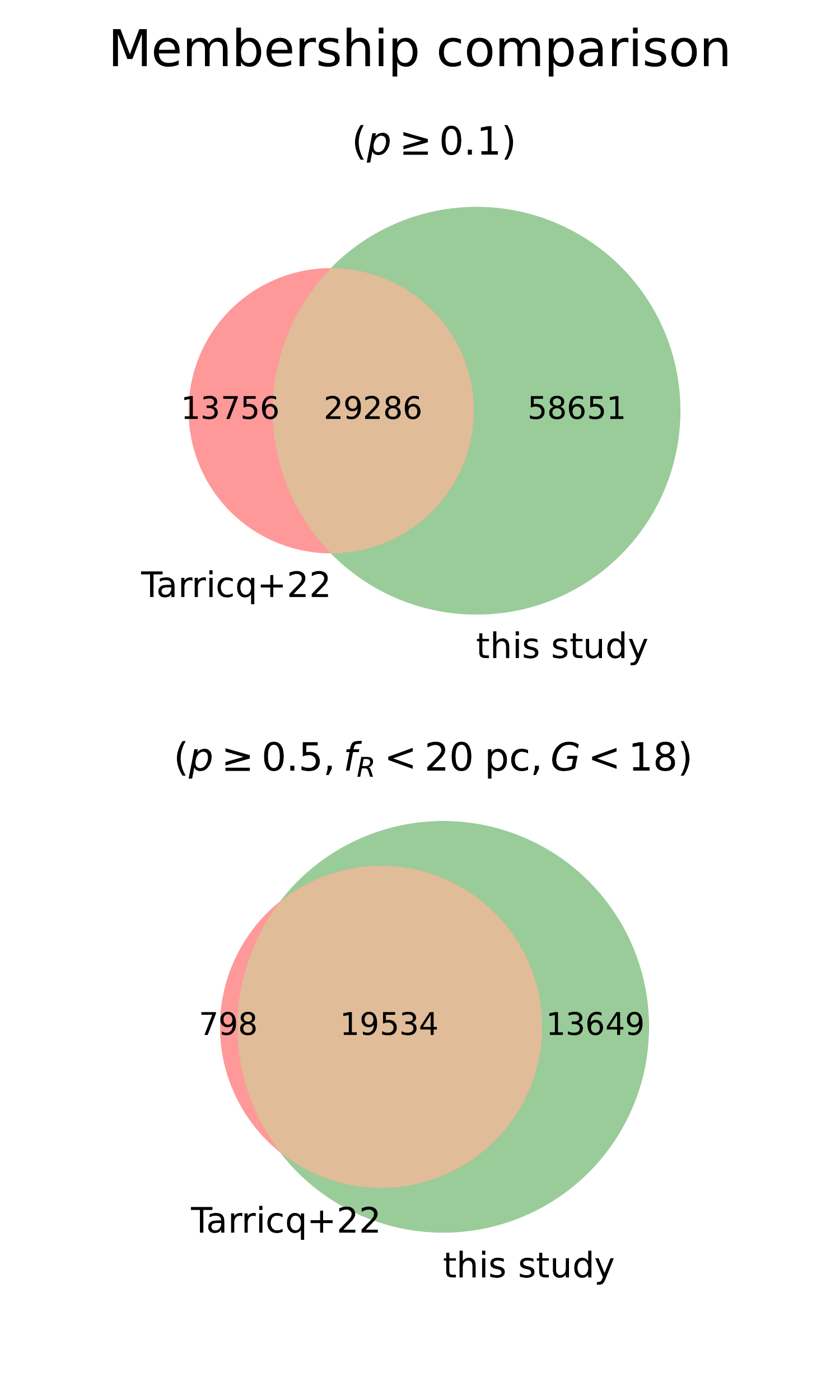}
\caption{Venn diagrams comparing the combined membership lists of the 167 test OCs from T22 and this study. The top figure compares the members with a membership probability of $p \geq 0.1$, while the bottom figure compares the members with a membership probability $p \geq 0.5$, a projected radius of less than 20 pc and a $G$-magnitude brighter than 18. The number of members that only occur in T22 are labelled in red, the members only in this study in green and the overlap in orange.}
\label{fig:member_comparison}
\end{figure}

In Fig.~\ref{fig:member_comparison}, we present two Venn diagrams that show the overlap between the members from T22 and the members in this study. The top figure in Fig.~\ref{fig:member_comparison} includes all members with a membership probability $p \geq 0.1$ and shows that we generally find the majority of the T22 members and also a significant amount of additional members. In the most extreme cases, over 90\% of the members we obtain for a single cluster are not in the corresponding T22 membership list. In the subsequent sections, we discuss the origins of the differences in the membership lists. 

\begin{figure*}
\centering
\includegraphics[width = 1.\textwidth]{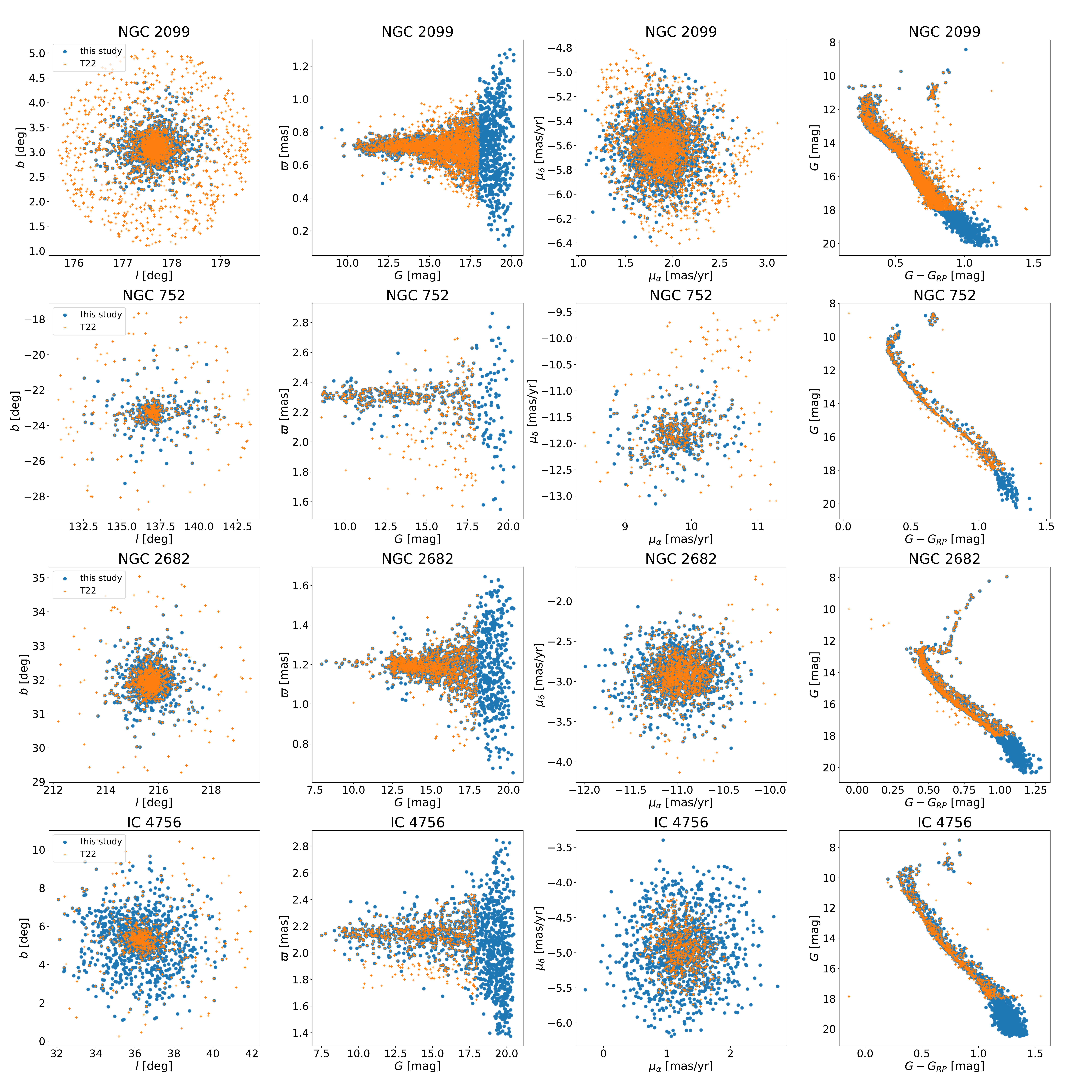}
\caption{Distributions of $p \geq 0.1$ members of NGC 2099, NGC 752, NGC 2682 and IC 4756 found in this study (blue) and by T22 (orange) in (from left to right) the sky position, parallax, proper motion and CMD.}
\label{fig:member_distributions}
\end{figure*}

In Fig. \ref{fig:member_distributions}, we compare the member distributions in sky position, proper motion, parallax and the CMD of four OCs: NGC 2099, NGC 752, NGC 2682 and IC 4756. These plots serve as examples of the member distributions we obtain and will be used as a reference to highlight some of the trends we observe when comparing the membership lists. 

\subsection{Projected radius and $G$-magnitude}

As both methods determine OC membership in a different way, the discrepancy between the membership lists is to be expected to some degree, however, many members are excluded from either list for trivial reasons. For example, in contrast to our candidates, T22 a priori excluded sources with $G > 18$ from their membership list. On the other hand, our method ascribes lower membership probabilities to sources with large projected radii, while the T22 membership probability does not depend on the sky position. If we analyse the members from T22 that we missed, i.e. the members from T22 we either select as non-members or ascribe a membership probability of $p < 0.1$, which together make up 32\% of the total number of T22 members, we find that 73\% of these were selected as candidate, but that the average projected radius of these candidates is 38.3 pc with a standard deviation of 3.1 pc. Sources beyond this radius are typically given very low membership probabilities as the training members from \cite{2020A&A...640A...1C} usually do not extend far beyond the core of the cluster. A clear example of this can be seen in the sky position plot of NGC 2099 in Fig.~\ref{fig:member_distributions}, where the outskirts are only populated by T22 members. In order to show the significance of these differences, we present a similar comparison in the bottom plot of Fig.~\ref{fig:member_comparison} where we only consider sources with $G<18$ and with projected radii of less than 20 pc. As the high-probability sources are more relevant for comparison than the low-probability sources, we also consider only sources with membership probability $p \geq 0.5$ for this plot. After these cuts, a total of $33\,184$ (38\%) members in our study and $20\,332$ (47\%) T22 members remain. This comparison shows that we find nearly all of the probable ($p \geq 0.5$) T22 members within a 20 pc radius. For 61 OCs, we find 100\% of these T22 members. We can also see that the fraction of new members is generally lower, as a large proportion of all $p \geq 0.1$ members we obtain are $G > 18$ members, which are excluded from the bottom Venn diagram. The median fraction of $p \geq 0.1$ members we obtain with $G > 18$ is 43.5\%. For nearby OCs, which have more faint sources with high probabilities due to smaller astrometric uncertainties, the fraction of members we obtain with $G>18$ and $p \geq 0.1$ can be as large as 70-80\%.  

\subsection{Parallax and proper motion}

The remaining differences are primarily the result of the different treatments of the parallax and proper motion dimensions, which comprise the data utilised by both methods. If we consider only the sources used for the bottom plot in Fig.~\ref{fig:member_comparison}, i.e. sources with $p \geq 0.5$, $f_{r} < 20 \; \mathrm{pc}$, $G < 18$, we obtain median parallax and proper motion features $f_{\varpi}=-0.003^{+0.067}_{-0.076}$ and $f_{\mu}=0.20^{+0.30}_{-0.13}$ for our members, where the bounds indicate the 15th and 85th percentile, while the same statistics for the selected T22 members are $f_{\varpi}=-0.002^{+0.047}_{-0.052}$ and $f_{\mu}=0.16^{+0.41}_{-0.09}$. Thus our method is, on average, effectively less 'strict' in the parallax and proper motion dimension. In contrast with this trend, some OCs have T22 member distributions that are much more extended than the corresponding distributions of the training members. For example, the OCs UPK 303, COIN-Gaia 30, ASCC 58, NGC 1901, and COIN-Gaia 13 have a much broader distribution of T22 members in proper motion compared to the training members, resulting in many of T22 members to be selected as non-members by our method. In Fig.~\ref{fig:member_distributions}, NGC 752 is another example of this. The statistics of all T22 members which were not selected as candidates also show the relative strictness of our proper motion condition. Of these missing T22 members, 67\% failed the proper motion condition. By comparison, only 27\% failed the parallax condition and only 20\% failed the isochrone condition. Clear examples of T22 members excluded by the parallax condition can be seen in the parallax plot of cluster IC 4756 in Fig.~\ref{fig:member_distributions} and examples for T22 members excluded by the isochrone condition can be seen for the clusters NGC 2682 and NGC 2099.

\section{Summary and Conclusions}
\label{sec:conclusions}

We have developed a methodology to find new OC members in {\em Gaia} DR3 for the population of known OCs. This methodology is based on a deep neural network architecture, which is able to learn the distribution of highly reliable OC members in a high dimensional space, meaning five-dimensional astrometry and photometry, and retrieve new members based on the similarities in these parameters. To train our method, we take advantage of the high-quality OC catalogue built using {\em Gaia} DR2 \citep{2020A&A...640A...1C} and EDR3 \citep{2022A&A...661A.118C}, which contains around $2\,500$ OCs with membership lists, mean astrometric parameters and astrophysical information.

The method is available as an open-source python tool under \texttt{\url{https://github.com/MGJvanGroeningen/gaia_oc_amd}}. This python package has built-in functions to go through all the steps described in the previous sections, from querying OC members and their mean parameters, generating the different cone searches in the {\em Gaia} archive and creating the member, non-member and candidates datasets, to training the model and using it to find new OC members. Documentation and a step-by-step tutorial in the form of a python notebook are included within the package.

When comparing our results with independent membership determinations for a subset of the OC catalogue \citep{Tarricq_2022}, we are able to retrieve $100\%$ of their members within $20$ pc of the cluster centre while adding some new members at bright magnitudes $(G \leq 18)$. More importantly, we are able to extend membership lists to fainter magnitudes, down to the {\em Gaia} magnitude limit, in a homogeneous way for the first time on the whole OC catalogue. This is needed for forthcoming spectroscopic surveys such as WEAVE or 4MOST, whose input target lists are fully based on {\em Gaia}, and in their low-resolution modes, they can observe sources fainter than $G = 18$ mag. These surveys will complement {\em Gaia} with radial velocities for stars with $G_{\text{RVS}} \sim 16$ and chemical abundances for all the observed stars. In the context of this work, this will allow us to further refine OC membership lists and retrain our method for a more accurate membership determination. 

Having more complete membership lists for the OCs also enables further scientific applications. So far, {\em Gaia} has redefined the OC census in terms of a better characterisation of their astrometric properties, the addition of hundreds of new objects to the catalogue or the estimation of some astrophysical properties which only depend on the shape of the OC isochrone in the CMD. However, further improvements to the OC catalogue such as the estimation of masses or the dynamical evolution of OCs (and their members) through the Galactic disc, rely on a complete description of the OC in the whole {\em Gaia} magnitude range and the distribution of its member stars in the CMD, also accounting for possible selection effects on these stars \citep{2022arXiv220809335C}.

\begin{acknowledgements}

This work has made use of results from the European Space Agency (ESA)
space mission {\it Gaia}, the data from which were processed by the {\it Gaia
Data Processing and Analysis Consortium} (DPAC).  Funding for the DPAC
has been provided by national institutions, in particular, the
institutions participating in the {\it Gaia} Multilateral Agreement. The
{\it Gaia} mission website is \url{http: //www.cosmos.esa.int/gaia}. The
authors are current or past members of the ESA {\it Gaia} mission team and
of the {\it Gaia} DPAC. This research has made use of the tool provided by \Gaia DPAC (\url{https://www.cosmos.esa.int/web/gaia/dr3-software-tools}) to reproduce (E)DR3 \Gaia photometric uncertainties described in the GAIA-C5-TN-UB-JMC-031 technical note using data in \cite{Riello_2021}. This work was (partially) funded by the Spanish MICIN/AEI/10.13039/501100011033 and by "ERDF A way of making Europe" by the “European Union” through grants RTI2018-095076-B-C21 and PID2021-122842OB-C21, and the Institute of Cosmos Sciences University of Barcelona (ICCUB, Unidad de Excelencia ’Mar\'{\i}a de Maeztu’) through grant CEX2019-000918-M.

\end{acknowledgements}

\bibliographystyle{aa} 
\bibliography{bibliography}

\begin{appendix}
\section{Model architecture}\label{app:model_architecture}

A diagram of the complete DS model is given in Fig.~\ref{fig:model}. The first part of the model consists of a Permutation Equivariant Layer (PEL) and an Exponential Linear Unit (ELU) \citep{2015arXiv151107289C} activation function , which is repeated five\footnote{In the original model from \cite{Zaheer_2017}, this block is repeated only three times. This is the only difference compared to the version from \cite{Oladosu_2020} and thus compared to our model as well.} times. 

\begin{figure}[h!]
    \centering
    \includegraphics[width=0.8\columnwidth]{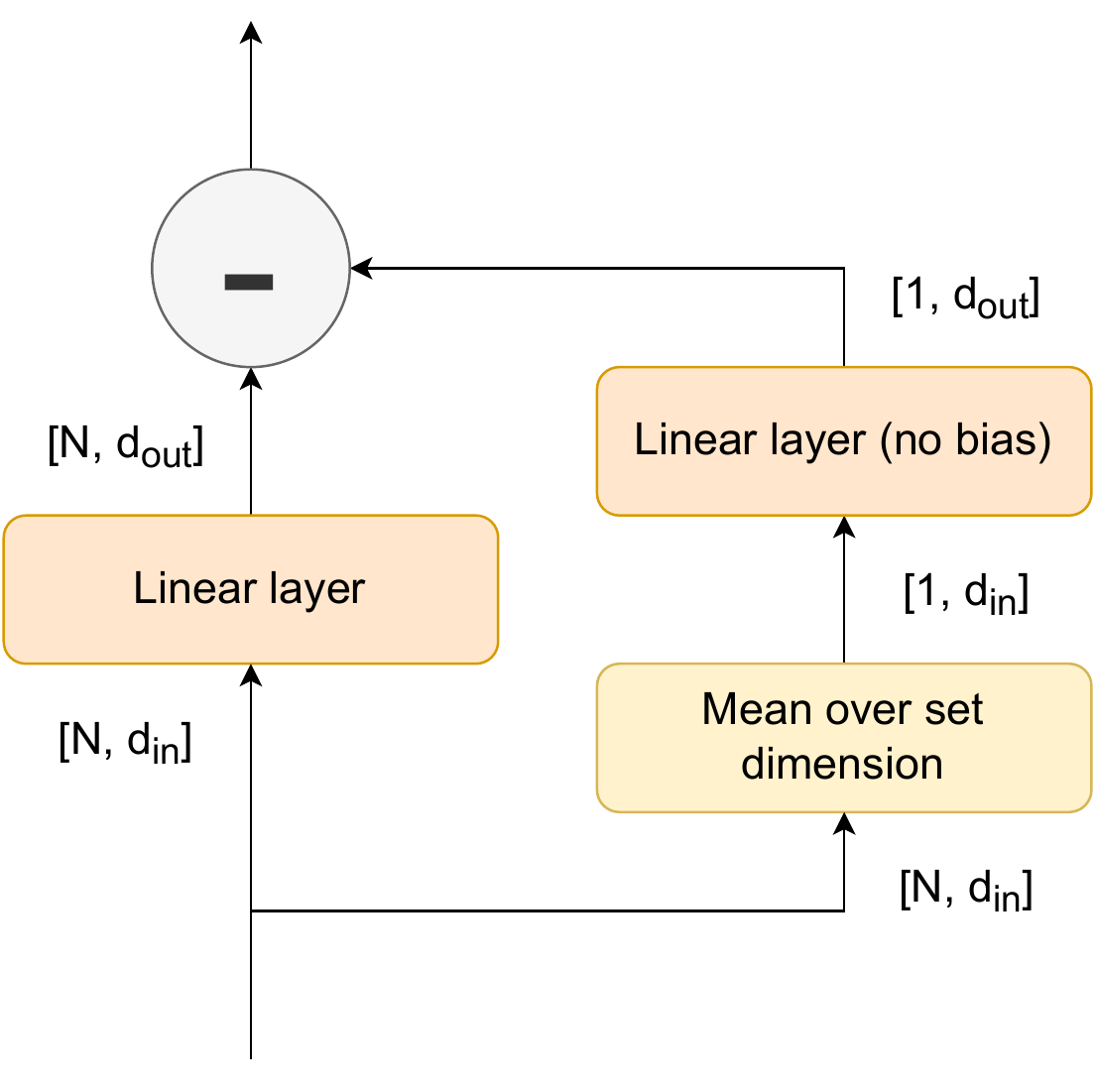}
    \caption{Diagram containing the operations in the permutation equivariant layer. The variables in the brackets indicate the dimensions of the tensors between operations. The batch dimension is left out for clarity. In the linear layer on the right track, the biases are set to zero.}
    \label{fig:perm_equi_layer}
\end{figure}

In the PEL, expanded into its components in Fig.~\ref{fig:perm_equi_layer}, the input follows two parallel tracks. In Fig.~\ref{fig:perm_equi_layer}, the track on the left contains one linear layer, which performs the operation

\begin{equation}
    \bm{x}^{\prime} = \bm{W}_{n} \bm{x} + \bm{b}_{n},
\end{equation}
on its input vector $\bm{x}$, with dimensionality $d_{\mathrm{in}}$, and returns a new vector $\bm{x}^{\prime}$, with dimensionality $d_{\mathrm{out}}$. The weight matrix $\bm{W}_{n}$ and bias vector $\bm{b}_{n}$ of linear layer $n$ constitute trainable parameters of the model, which are to be optimised during the training process. In the right track, the mean over the set dimension is taken first, followed by another linear layer. Finally, the output from the right track is subtracted from the output of the left track. Note that, as the input to the first PEL consists of the training features, the mean features of the support set members are part of the result from the first 'mean over set dimension'. The membership prediction is thus partly based on the mean features of the members in the support set. The term 'permutation equivariant' refers to the feature of the PEL that a permutation (of the set dimension) of the input gives the same result as the same permutation on the output

\begin{equation}
    \mathrm{PEL}(\mathrm{permutation}(\bm{X})) = \mathrm{permutation}(\mathrm{PEL}(\bm{X})). 
\end{equation}

After the PEL blocks, taking the mean over the set dimension guarantees the invariance of the output with respect to a permutation of the model input, fulfilling the precondition for a model operating on sets. This is followed by a dropout layer, which randomly sets elements of the input tensor to zero during training, with a 50\% probability for each element. This prevents over-reliance on certain features of the input which helps prevent overfitting to the training data \citep{Hinton_2012}. The final linear layer transforms its input, which is a vector with hidden dimension $d_{\mathrm{h}}$, to a 2-dimensional vector, corresponding to the two classes: member and non-member. The softmax layer then converts values in the 2-dimensional vector to values that sum to 1 and can thus be interpreted as a probability for each class. Finally, the class with the highest probability is attributed to the candidate member included in the model input.

\begin{figure}[h!]
    \centering
    \includegraphics[width=0.8\columnwidth]{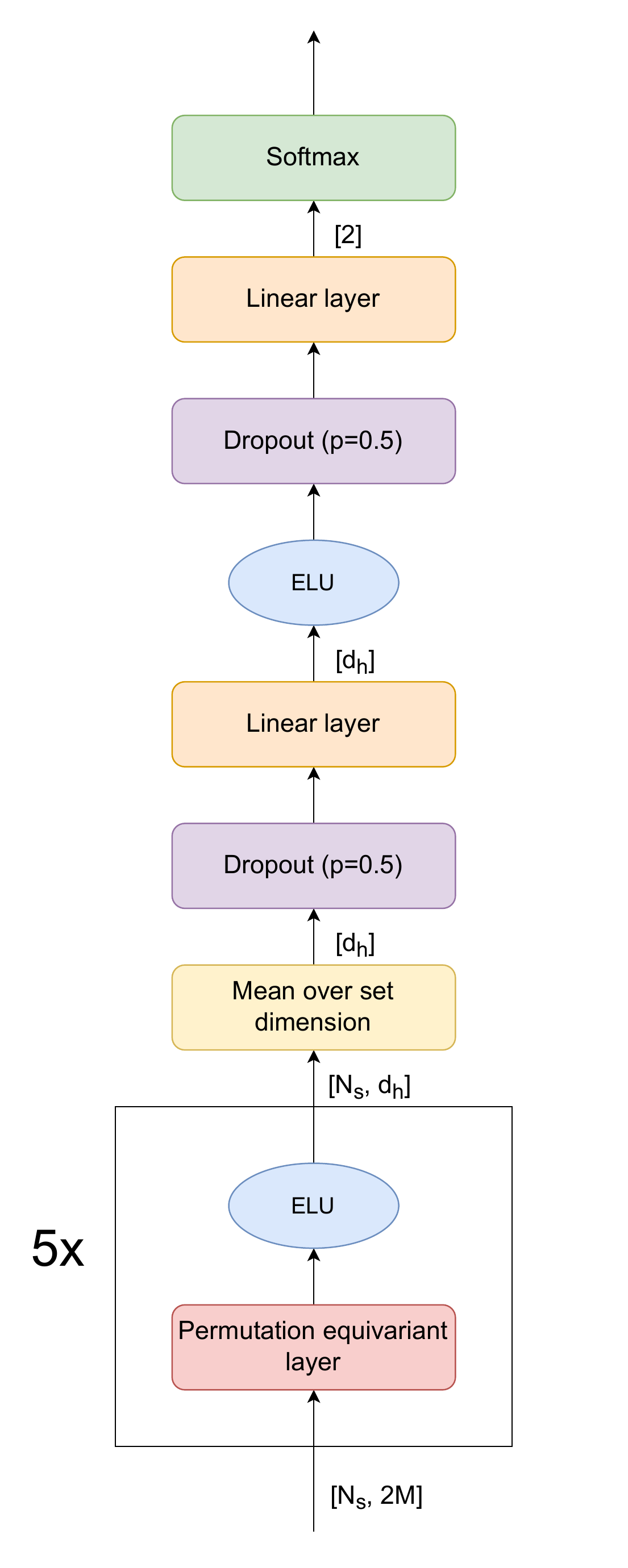}
    \caption{A diagram of the complete DS model. Details of the permutation equivariant layer are given in Fig.~\ref{fig:perm_equi_layer}. The variables and values in the brackets indicate the dimensions of the tensors between operations and the batch dimension is left out for clarity. The symbol $N_{s}$ refers to the size of the support set, $M$ to the number of training features and $d_{h}$ to the hidden dimension of the network.}
    \label{fig:model}
\end{figure}

\label{app:appendix_A}

\end{appendix}
\end{document}